\documentclass[preprintnumbers, floatfix,letterpaper, nofootinbib, twocolumn,aps,prd,epsfig]{revtex4-1}
\usepackage{bm,graphicx,dcolumn,epstopdf,epsf, latexsym,mathbbol, amssymb,amsmath,color,slashed, mathrsfs,mathcomp,simplewick}
\usepackage[center]{subfigure}
\usepackage{multirow}
\usepackage{makecell}
\usepackage[colorlinks,linkcolor=blue,citecolor=blue,urlcolor=blue]{hyperref}
\usepackage[center]{subfigure}

\begin{document}
\allowdisplaybreaks
 \newcommand{\bq}{\begin{equation}}
 \newcommand{\eq}{\end{equation}}
 \newcommand{\bqn}{\begin{eqnarray}}
 \newcommand{\eqn}{\end{eqnarray}}
 \newcommand{\ban}{\begin{align}}
 \newcommand{\ean}{\end{align}}
  \newcommand{\nb}{\nonumber}
 \newcommand{\lb}{\label}
 \newcommand{\f}{\frac}
 \newcommand{\p}{\partial}
\newcommand{\PRL}{Phys. Rev. Lett.}
\newcommand{\PLB}{Phys. Lett. B}
\newcommand{\PRD}{Phys. Rev. D}
\newcommand{\CQG}{Class. Quantum Grav.}
\newcommand{\JCAP}{J. Cosmol. Astropart. Phys.}
\newcommand{\JHEP}{J. High. Energy. Phys.}
\newcommand{\NPB}{Nucl. Phys. B}
\newcommand{\Doi}{https://doi.org}
\newcommand{\red}{\textcolor{black}}
\newcommand{\magenta}{\textcolor{magenta}}
\newcommand{\gre}{\textcolor{green}}
\title{Gravitational axial perturbations of Schwarzschild-like black holes in dark matter halos}

\author{Chao Zhang ${}^{a, b, c}$}
\email{chao123@zjut.edu.cn}

\author{Tao Zhu${}^{a, b}$}
\email{corresponding author:  zhut05@zjut.edu.cn}

\author{Anzhong Wang${}^{d}$}
 \email{anzhong$\_$wang@baylor.edu}

\affiliation{
${}^{a}$ Institute for Theoretical Physics and Cosmology, Zhejiang University of Technology, Hangzhou, 310032, China\\
${}^{b}$ United Center for Gravitational Wave Physics (UCGWP), Zhejiang University of Technology, Hangzhou, 310032, China\\
${}^{c}$ College of Information Engineering, Zhejiang University of Technology, Hangzhou, 310032, China \\
${}^{d}$ GCAP-CASPER, Physics Department, Baylor University, Waco, Texas 76798-7316, USA}

\date{\today}

\begin{abstract}

Gravitational waves emitted by distorted black holes, such as those arising from the coalescence of binary black holes, or a falling compact star into a supermassive black hole, carry not only information about the corresponding spacetime but also the information about the environment surrounding the black holes. In this paper, we study the effects of the dark matter halos with three different density profiles on the gravitational axial perturbations of a Schwarzschild-like black hole. For this purpose, we first consider modified Schwarzschild black holes with three different dark matter profiles and derive the equation of motion of the axial perturbations of the modified Schwarzschild metric. It is shown that by ignoring the dark matter perturbations,  a Regge-Wheeler-like master equation with a modified potential for the axial perturbation can be obtained explicitly. Then we calculate the complex frequencies of the quasinormal modes of the Schwarzschild-like black hole in the dark matter halos by applying the sixth-order WKB method. The corresponding gravitational wave spectra with the effects of the dark matter halos have also been discussed.

\end{abstract}

\maketitle

\section{Introduction}
\renewcommand{\theequation}{1.\arabic{equation}} \setcounter{equation}{0}

Black holes (BHs) are one of the most mysterious phenomena in the Universe. The existence of BHs provides us a perfect way to test gravitational effects under extremely strong gravity, such as the formation of gigantic jets of particles and disruption of neighboring stars \cite{Kimet2019}. On the other hand, from the theoretical point of view, BHs are also unique labs to test modified theories of gravity that are different from general relativity (GR) (see, e.g., \cite{Xiang2019, Chao2020, Chao2020b}). 

The outbreak of interest on BHs has further gained momenta after the detection of the shadow of the M87 central BH  with the Event Horizon Telescope (EHT) \cite{EHTL1, EHTL2, EHTL3, EHTL4, EHTL5, EHTL6, EHTL7, EHTL8}. In fact, it is widely believed that the central region of many galaxies contain supermassive BHs. The shadow image of the M87 central BH captured by EHT is in good agreement with the prediction of the spacetime geometry of a BH described by the Kerr metric \cite{Kerr1963}. Nonetheless, since it is believed that up to $ 90\%$ of the matter in a host galaxy is made up by dark matter (DM) \cite{Kimet2020}, it is natural to expect that the DM halo surrounding a central BH will bring small deviations to the Kerr metric. 

Although DM has not been directly detected, many evidences imply its existence. In addition, because of the rotation curves of galaxies, the mass-light ratio of elliptical galaxies, the cosmic microwave background radiation, and the large-scale structure of the cosmos, a lot of research has been done on DM and proposed various DM models \cite{Zhaoyi2020}. Of course, there are many open questions related to DM, for instance, the spatial density distribution of DM in galaxies, solutions of Einstein's field equations when considering the DM background, and more.

In this paper, we focus on the effects of DM halos on gravitational waves (GWs) emitted from the central BH located in a galaxy. The detection of the first GW from the coalescence of two massive BHs by advanced LIGO marked the beginning of a new era, the GW astronomy \cite{Ref1}. Following this observation, about 90 GW events have been identified by the LIGO/Virgo/KAGRA scientific collaborations (see, e.g., \cite{GWs,GWs19a,GWs19b, GWsO3b}). In the future, more ground- and space-based GW detectors will be constructed \cite{Moore2015, Aso2013, Liu2020, Taiji2}, which will enable us to probe signals with a wider frequency band. This triggered the interests on the quasinormal modes (QNMs) from GWs, including those from the late-merger and ringdown stages of coalescence \cite{Berti18} as well as those from supermassive BHs, e.g, central BHs in a galaxy. \red{The detection of QNMs from the ringdown stage will provide a unique way to probe the matter environment surrounding a BH.} For example, it is shown that \red{the shift from GR on the QNMs due to the surrounding ultralight bosons might be detectable in the future by the observational data }from LISA-like missions \cite{Chung:2021roh}. The effects of the dark matter or matter distributing around BHs on the QNMs have also been considered in \cite{Cardoso:2021wlq, Liu:2021xfb, Konoplya}.

From the theoretical point of view, QNMs are eigenmodes of dissipative systems. The information contained in QNMs provide the keys for revealing whether BHs are ubiquitous in our Universe, and more important whether GR is the correct theory to describe the event even in the strong field regime. Readers may find more details in \cite{Berti2009}. Basically, the QNM frequency $\omega$ contains two parts, the real part and the imaginary part. Its real part gives the frequency of vibration while its imaginary part provides the damping time. In other words, the frequency we are going to calculate is a complex number (although it could be purely imaginary in certain circumstances).

In fact, in GR according to the no-hair theorem, an isolated BH is completely characterized by only three quantities, mass, spin angular momentum and electric charge. Astrophysically, we expect BHs to be neutral, so they ought to be described by the Kerr solution. Then, the quasinormal frequencies and damping times will depend only on the mass and angular momentum of a BH. Therefore, to extract the physics from QNMs, at least two QNM signals are needed. This will require the signal-to-noise ratio (SNR) to be of the order 100. Although such high SNRs are not achievable right now, it has been shown that they may be achievable once the advanced LIGO and Virgo reach their designed sensitivities. In any case, it is certain that they will be detected by the ground-based third-generation detectors, such as Cosmic Explorer or the Einstein Telescope, as well as the space-based detectors, including LISA, TianQin \cite{Shi2019}, Taiji \cite{Taiji2}, and DECIGO \cite{Moore2015}.

For this purpose, we first consider modified Schwarzschild black holes with three different DM profiles and derive the equation of motion of the axial perturbations \footnote{{Normally, the GW spectrum of QNMs can be studied via black hole perturbations. In the background spacetime which is static and spherically symmetric, the metric perturbations can decouple into two independent parts, the polar one and the axial one. In this paper we only consider the latter case. As we discuss later, the axial perturbation considered in this paper is equivalent to the odd-parity perturbation, which corresponds to the parity of $(-1)^l$, with $l$ being an index inherited from spherical harmonics \cite{Cardoso2001}.}} of the modified Schwarzschild metric. It is shown that by ignoring the DM perturbations,  a Regge-Wheeler (RW)-like master equation with a modified potential for the axial perturbation can be obtained explicitly. Then we calculate the frequencies of the QNMs of the Schwarzschild-like black hole in the DM halos by applying the sixth-order WKB method. The corresponding GW spectra with the effects of the DM halos have also been discussed.

Here, we consider the QNMs of the axial metric perturbations of a Schwarzschild-like BH surrounded by the DM halos. Several different background metrics are investigated by considering three different DM profiles. These metrics for different dark matter halo profiles can be found in \cite{Kimet2019, xu_JCAP}. Note that recently the metric of a black hole immersed in dark matter spike has also been derived \cite{Xu:2021dkv}. In addition, the Sgr $\text{A}^\ast$ black hole (located in the center of Milky Way Galaxy) and the M87 galactic central black hole are what we focus on. In other words, the structure we consider is a BH located at the center of a galaxy. By comparing the resultant $\omega$'s with their counterparts for the Schwarzschild case, we see the influence of DM halo on QNMs as well as GWs from central BHs. At the same time, we will also estimate the detectability of these deviations from the Schwarzschild case. Finally, by varying some profile-dependent constants, one can find out how these $\omega$'s are deviating from the Schwarzschild case with the effects of the DM halos. 

The rest of this paper is organized as follows. Sec. II shows some basic information of the three density profiles of DM halos that we investigate for the calculations of QNMs. After that, in Sec. III we show how to derive a Regge-Wheeler-like master equation from the axial perturbation and Einstein's field equations. At the same time, we show a brief analysis of the effective potentials under different DM profiles (including that from the Schwarzschild case). Section IV contains two parts. In the first part, we present some resultant $\omega$'s. Some concluding remarks are given by comparing them with their counterparts in the Schwarzschild case. For the second part, we test the effects of model-dependent constants. Some results are shown together with an analysis. Finally, Sec. V provides our main conclusions as well as some outlooks for future work. 

Through out the paper, we adopt the unit system so that $c=G_N=1$, where $c$ is the speed of light, while $G_N$ stands for the  gravitational constant observed on Earth. In this way, we still have one degree of freedom to choose the unit for length. This is done later by setting ${r_{\rm MH}}=1$, where {$r_{\rm MH}$} means the radius of the metric horizon (MH) of the BH that we are focusing on. {In this paper, all the greek indices run from 0 to 3. Other usage of indices is indicated explicitly when it is necessary.} 

\section{Black hole solutions in dark matter halo}
\renewcommand{\theequation}{2.\arabic{equation}} \setcounter{equation}{0}

Normally, the black hole spacetimes are not clean and are affected by the surrounding matters. In this section, we consider the spherically symmetric static black hole solutions with several different DM halo profiles. In general, a Schwarzshild  BH in the DM halo is described by the metric \cite{xu_JCAP}
\bqn\lb{metric_dark}
ds^2 & = & - G(r) dt^2 + \frac{1}{G(r)} dr^2 + r^2 d\theta^2 + r^2 \sin^2\theta d\varphi^2, \nb\\
\eqn
where $G(r)$ denotes the function that describes the effects of the DM halos and BH on the metric and it reduces to the Schwarzschild solution, viz., $G(r)=1- 2M/r$ \cite{CarrollB}, when the DM is absent. For different profiles of the DM halo, the function $G(r)$ is different \cite{xu_JCAP}. In the following, we present the function $G(r)$ for each profile individually.

\subsection{URC profile}

In the universal rotation curve (URC) profile of the DM halo, the distribution of the DM is described by \cite{URC1} (see also \cite{halo_review} for a review)
\bqn
\rho(r) = \frac{\rho_0 r_0^3}{(r+r_0)(r^2+r_0^2)},
\eqn
where $\rho_0$ is the central density and $r_0$ is the characteristic radius of the DM halo. According to the observations on the M87 galaxy, the best fit values for the parameters of the URC profile are $\rho_0 = 6.9\times 10^6 \text{M}_{\odot}/{\rm kpc}^{3}$ and $r_0 = 91.2\; {\rm kpc}$ \cite{Salucci_M87}. While in the Milky Way Galaxy, we have $\rho_0 = 5.2 \times 10^7 \text{M}_{\odot}/{\rm kpc}^{3}$ and $r_0 = 7.8\; {\rm kpc}$ \cite{dark_matter}. With this halo profile, the function $G(r)$ in the metric (\ref{metric_dark}) is given by \cite{Kimet2019, kimet_shadow}
\bqn
\lb{URC_Gr}
G(r) &=& e^{-2 \pi ^2 \rho_0 r_0^2} \left(1+\frac{r^2}{r_0^2}\right)^{-\frac{2 \,\rho_0 r_0^3 \pi}{r}(1-\frac{r}{r_0})} \nb\\
&& \times \left(1+\frac{r}{r_0} \right)^{-\frac{4\, \rho_0 r_0^3  \pi}{r}(1+\frac{r}{r_0})} \nb\\
&& \times \exp\left[\frac{4\, \rho_0 r_0^3 \pi \arctan(\frac{r}{r_0})(1+\frac{r}{r_0})}{r}   \right]- \dfrac{2M}{r}.\nb\\
\eqn
Here $M = 6.5 \times 10^9~\text{M}_{\odot}$ for the M87 central BH and $M = 4.3 \times 10^6~\text{M}_{\odot}$ for the Sgr $\text{A}^\ast$ BH.

\subsection{The CDM halo with NFW profile}

The cold dark matter (CDM) halo with a Navarro-Frenk-White (NFW) profile is obtained by $N$-body simulations, which has a universal spherically averaged density profile \cite{NFW, Kimet2019}
\bqn
\rho(r) = \frac{\rho_0}{(r/r_0)(1+r/r_0)^2},
\eqn
where $\rho_0$ is the density of the Universe at the moment when the halo collapsed and $r_0$ is the characteristic radius.  According to the observations on the Milky Way Galaxy \cite{dark_matter}, the best fit values for the parameters $\rho_0$ and $r_0$ for the NFW profile are $\rho_0 = 5.23 \times 10^7 \text{M}_{\odot}/{\rm kpc}^{3}$ and $r_0=8.1 \; {\rm kpc}$. On the other hand, for the M87 galaxy, we have  $\rho_0 = 0.008 \times 10^{7.5}~\text{M}_{\odot}/ \text{kpc}^3$ (see \cite{Oldham2016}) and ${r_0} = 130~\text{kpc}$ \cite{Kimet2019}. With this halo profile, the function $G(r)$ in the metric (\ref{metric_dark}) is given by \cite{xu_JCAP}
\bqn
\lb{Gr2}
G(r) = \left(1+\frac{r}{{r_0}}\right)^{-\frac{8 \pi G_N \rho_0 r_0^3}{c^2 r}} - \frac{2 G_N M}{c^2 r}.
\eqn
Here $M = 4.3 \times 10^6~\text{M}_{\odot}$ is the mass of Sgr $\text{A}^\ast$ BH, and $M = 6.5 \times 10^9~\text{M}_{\odot}$ is the mass of M87 central BH.

\subsection{The SFDM model}

For the scalar field dark matter (SFDM) model \cite{Xian2018}, the energy density profile for a DM halo is given by
\bqn
\rho(r) = \frac{\rho_0 \sin(\pi r/r_0)}{\pi r /r_0},
\eqn
where $\rho_0$ is the central density and $r_0$ is the radius at which the pressure and density are zero. In the Milky Way Galaxy, we have $\rho_0 = 3.43\times 10^7 \text{M}_{\odot}/{\rm kpc}^{3}$ and $r_0 = 15.7 \; {\rm kpc}$ \cite{Xian2018}. With this halo profile, the function $G(r)$ in the metric (\ref{metric_dark}) is given by
\bqn
\lb{Gr3}
G(r) = \exp\left[-\frac{8 G_N \rho_0 R^2}{\pi} \frac{\sin(\pi r/r_0)}{\pi r/r_0}\right] - \frac{2 G_N M}{c^2 r}. \nb\\
\eqn
Here $M = 4.3 \times 10^6~\text{M}_{\odot}$ is the mass of Sgr $\text{A}^\ast$ BH.

 \section{Regge-Wheeler-like equation for axial metric perturbations}
\renewcommand{\theequation}{3.\arabic{equation}} \setcounter{equation}{0}

In this section, we consider the linear gravitational perturbations $h_{\mu\nu}$ around Schwarzshild-like solutions. Let us first start with a general form of a spherically symmetric spacetime, given by \cite{Cruz2019}
\bqn
\lb{backg}
d s^2 &=& -G(r) dt^2 + F^{-1}(r) d r^2+H(r) d\Omega^2,
\eqn
where
\bqn
\lb{dOmega}
d \Omega^2 &=& d \theta^2+\sin^2\theta d \phi^2.
\eqn
Of course, for our case [cf. \eqref{metric_dark}], we have $G=F$ and $H=r^2$. For metric perturbations, let us start by describing the geometry of a linearly perturbed spherically symmetric background $\bar g_{\mu\nu}$
\bqn
\lb{pertb1}
{g_{\mu\nu} = \bar g_{\mu\nu} + h_{\mu\nu}, }
\eqn
where
\bqn
\bar g_{\mu\nu} = {\rm diag}\left(- G(r),\frac{1}{F(r)}, H(r), H(r)\sin^2\theta\right),~~
\eqn
where $h_{\mu\nu}$ denotes the linear perturbations of the background metric $\bar g_{\mu\nu}$. In general, the perturbation $h_{\mu\nu}$ can be split into pieces that transform as scalars, vectors, and tensors with respect to the symmetry of the spacetime. However, in two-dimensional maximally symmetric space $S^2$, it can be shown that the tensor perturbations with transverse-traceless conditions are identically zero \cite{cai_generalized_2013, takahashi_hawking_2010, takahashi_master_2010}. Thus, the metric perturbations can be split as scalar and vector perturbations, i.e., $h_{\mu\nu} =h^{\rm S}_{\mu \nu} + h^{\rm V}_{\mu\nu}$. Here, we note that the scalar perturbation \red{is also called a polar-type perturbation} (or even-parity perturbation), while the vector one is called an axial perturbation (or odd-parity perturbation) \cite{Berti2009}. In this paper, for simplicity, we only focus on the axial perturbations of the Schwarzshild-like solutions with different dark matter halos. 

We parametrize the axial perturbations in the form of \cite{Thomp2017}
\begin{widetext}
	\bqn
	\lb{hab}
	h_{\mu \nu} &=& \sum_{l=0}^{\infty} \sum_{m=-l}^{l} 
	\begin{pmatrix}
		0	& 0 & - C_{lm}\csc \theta \partial_\varphi & C_{lm} \sin \theta \partial_\theta\\
		0	& 0 & - J_{lm}\csc \theta \partial_\varphi & J_{lm} \sin \theta \partial_\theta\\
		sym	& sym & - W_{lm}\csc \theta \big(\cot \theta \partial_\varphi-\partial_\theta \partial_\varphi \big) &  sym \\
		sym	& sym & \frac{1}{2} W_{lm}\big(\csc \theta \partial^2_\varphi+\cos \theta \partial_\theta- \sin \theta \partial^2_\theta \big) &  W_{lm}\big(\cos \theta \partial_\theta- \sin \theta \partial_\theta \partial_\varphi \big)
	\end{pmatrix} 
	Y_{l m}(\theta, \varphi) \epsilon, 	\nb\\
	\eqn
\end{widetext}
where $C_{lm}$, $J_{lm}$, and $W_{lm}$ are functions of $t$ and $r$. The $Y_{l m}(\theta, \varphi)$ stands for the spherical harmonics \cite{Zettilib}, and the $l$ as well as $m$ in the index are integers. Here, $\epsilon$ is a real number, and \red{$|\epsilon| \ll 1$.}

From now on, we set  $m=0$ in \eqref{hab} so that $\partial_\varphi Y_{lm}(\theta, \varphi)=0$. As of now, the background has the spherical symmetry, and the corresponding linear perturbations do not depend on $m$   \cite{Regge57,Thomp2017}. In addition, by adopting the RW gauge \cite{Thomp2017}, we set $W_{lm}=0$.

The Einstein equations are given by
\bqn
\lb{Eab}
E_{\mu \nu} \equiv R_{\mu \nu}-\frac{1}{2} R g_{\mu \nu}= {8 \pi G_N T_{\mu\nu},}
\eqn
where $R_{\mu \nu}$ and $R$ are the Ricci tensor and Ricci scalar, respectively \cite{CarrollB}, and $T_{\mu\nu}$ denotes the energy-momentum tensor due to the presence of DM. For simplicity, here, we ignore the perturbation of the DM since its effects are expected to be negligible in comparing to the effects of DM from the modified background geometry. Then, expanding $E_{23}$ and $E_{13}$ to leading order of $\epsilon$, we obtain
\begin{widetext}
\bqn
\lb{PDE1}
0 &=& -\frac{{\dot C}}{G}+\frac{\left(G F' + F G'\right)}{2 G} J + F J', \\
\lb{PDE2}
0 &=& \frac{H' }{H} {\dot C}- {\dot C}' + {\ddot J} \nb\\
&& +\frac{1}{2}  \left[ F' G'+\frac{G \left \{H \left[F' H'+2 \left(F H''+l^2+l-2\right)\right]-F \left(H'\right)^2\right \}}{H^2}-\frac{F \left(G'\right)^2}{G}+F \left(2 G''+\frac{G' H'}{H}\right)\right] J,
\eqn
\end{widetext}
where a dot stands for the time derivative, while a prime stands for the derivative with respects to $r$. Notice that we have dropped the $lm$ in the subscripts of $C$ and $J$ for simplicity. After that, by combining \eqref{PDE1} and \eqref{PDE2}, we obtain the master equation
\bqn
\lb{master1}
\frac{d^2 \Psi(t, x)}{d x^2}-\left[\frac{d^2 }{d t^2} + V_{\rm eff}(r)\right] \Psi(t, x)  &=& 0,
\eqn
where
\bqn
\lb{Psi}
\Psi &\equiv& \left(\frac{H}{G F}\right)^{-1/2}J,\\
\lb{rast}
\frac{d r}{d x} &=& \sqrt{F G}, 
\eqn
and the effective potential is given by
\begin{widetext}
\bqn
\lb{Veff}
V_{\rm eff} &\equiv& \frac{1}{4} \left[2 F' G'+\frac{G \left \{H \left[F' H'+2 F H''+4 \left(l^2+l-2\right)\right]+F \left(H'\right)^2\right \}}{H^2}-\frac{2 F \left(G'\right)^2}{G}+F \left(4 G''+\frac{G' H'}{H}\right)\right].~~~~~
\eqn
\end{widetext}
By assuming $\Psi = e^{-i \omega t} \Psi(x)$, Eq.~\eqref{master1} could be written as
\bqn
\lb{master2}
\frac{d^2 \Psi(x)}{d x^2}+\left[\omega^2 - V_{\rm eff}(r)\right] \Psi(x) &=& 0.
\eqn

\begin{table*}
	\caption{Summary of the cases that we consider for the calculations of QNMs.}  
	\label{table0}   
\begin{tabular}{|c c c c c c c c c|} 
		\hline
		 & &  & &     & &  & & 
		\\[-7pt]
		Case & \quad  \quad & 	Galaxy & \quad  \quad & $G(r)$ & \quad \quad  & $\rho_0~\left(M_{\odot}/{\rm kpc}^{3} \right)$ & \quad \quad  & $r_0~({\rm kpc})$~~
		\\[1pt]
		\hline
		\hline
		 & &   & &     & &  & & 
		\\[-6pt]
		Schwarzschild & &  N/A   & &  $1-\frac{2 M}{r}$ & & $0$ & & N/A
		\\ [2pt] 
			\hline
		 & &   & &     & &  & & 
		\\[-7pt]
		Case 1 & &  M87  & &  \eqref{URC_Gr}  & & $6.9 \times 10^6$  & &  91.2
		\\ 
	      & &   Milky Way    & &  \eqref{URC_Gr}  & & $5.2 \times 10^7$  & &  7.8
		\\ [2pt] 
		\hline
				 & &   & &     & &  & & 
		\\[-7pt]
		Case 2 & &  M87  & & \eqref{Gr2}  & & $0.008 \times10^{7.5}$  & & 130
		\\ 
	      & &   Milky Way    & &  \eqref{Gr2}  & & $5.23 \times10^7$  & &  8.1
		\\ [2pt] 
    	\hline
				 & &   & &     & &  & & 
		\\[-7pt]
		Case 3 & &  Milky Way   & &  \eqref{Gr3}  & & $3.43 \times10^7$  & &  15.7
		\\ [2pt]
		\hline
	\end{tabular}
\end{table*}

Following Sec. II, we list the cases that we consider for the calculations of QNMs in Table \ref{table0} and provide some basic information for each case. They are referred to as case 1, 2 and 3, respectively. Note that the master equations and QNMs for case 2 and case 3 have also been studied in \cite{Liu:2021xfb} with relatively larger values of $r_0$ and $\rho_0$. The Schwarzschild case is also shown.

\begin{figure}[htb]
	\includegraphics[width=\columnwidth]{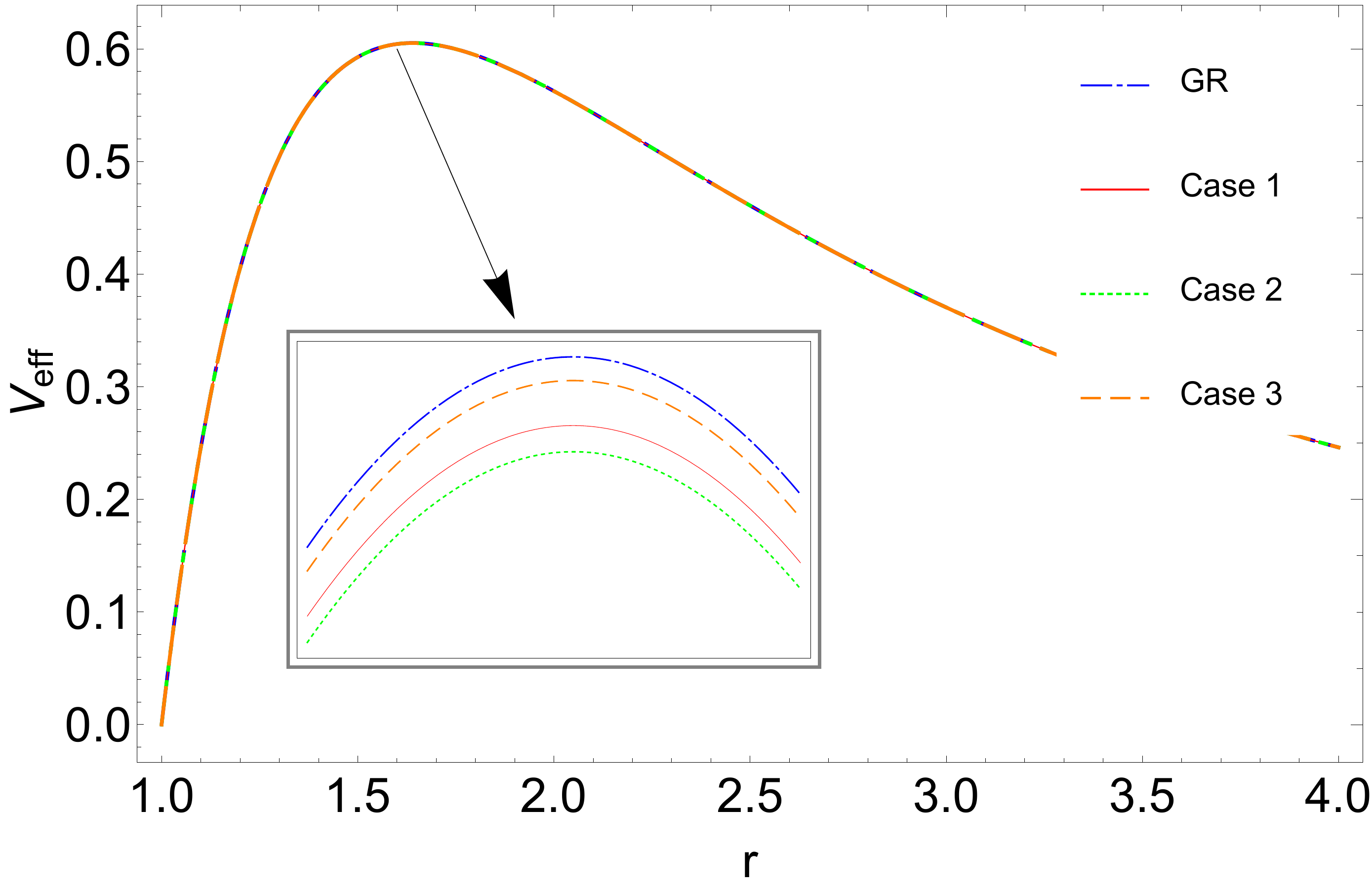} 
	\caption{ Behaviors of {$V_{\rm eff}$} for different cases listed in Table \ref{table0}, in which the data for the Milky Way is selected for plotting. In addition, we have chosen $l=2$. {Note that there is an inserted figure showing the amplification of the region around the stationary points of these curves.} Also note that here we are using the unit system so that $c=G_N={r_{\rm MH}}=1$.} 
	\label{plot1}
\end{figure} 

To find the difference between different cases, the $V_{\rm eff}$'s for each case listed in Table \ref{table0} are plotted in Fig. \ref{plot1}. To make them at the same starting line, the data for the Milky Way are selected for plotting. In addition, as an example, we have set $l=2$ for this plot. Notice that, as noted earlier, here we are using the unit system so that $c=G_N={r_{\rm MH}}=1$. From Fig. \ref{plot1}, it is very clear that the deviation on $V_{\rm eff}$'s between each two cases in Table \ref{table0} is quite small since these curves are almost overlapped. Actually, it is because of this, that an amplification of the region around the stationary points of these curves of $V_{\rm eff}$'s is inserted, so that readers can find more details. That implies we may obtain quite similar QNMs from these cases in the end. As we see, this is indeed the case. 

\section{QNMs of Schwarzshild-like black hole in DM halos}
\renewcommand{\theequation}{4.\arabic{equation}} \setcounter{equation}{0}

With the master equation given by \eqref{master2}, we are ready to solve for the corresponding $\omega$, viz., the QNM, for a specific set of $\{G, F, H\}$, including the cases listed in Table \ref{table0} (as well as their extensions by varying $\rho_0$ or $r_0$). Note that the QNMs for cases 2 and 3 have also been considered in \cite{Liu:2021xfb} with larger values of $\rho_0$ and $\rho_0$. Notice that, $\omega$, in general, is a complex number, often written as $\omega_{l m n}$ \cite{Berti2009}, where $l$ and $m$ are inherited from spherical harmonics while $n$ is the overtone index. However, since we have set $m=0$, it is left with two indices only, i.e., $l$ and $n$ (For simplicity, before stimulating any confusions, we write $\omega_{ln}$ as $\omega$).

Once again, for all the cases mentioned above, we adopt the unit system so that $c=G_N=r_{\rm MH}=1$. Here, the radius of the MH is given by $r_{\rm MH} = (2 G_N M)/c^2$ \cite{CarrollB}. In this way, the units for mass, time, and length are totally fixed.

\subsection{Calculations of QNMs for the cases in Table \ref{table0}}

So far, we have obtained the desired master equation [cf. \eqref{master2}], and we have several background metrics (see Table \ref{table0}). In addition, we also know the two boundary conditions, namely, the pure in-going wave at the MH and pure out-going wave at the spatial infinity \cite{Chandra1975}. With everything in hand, and given a set of $\{l, n\}$, we can find the corresponding $\omega_{ln}$, in principle.

QNMs in GR with the Schwarzschild case have been studied extensively. In this procedure, several different techniques of calculations were developed, for instance, the  Wentzel-Kramers-Brillouin (WKB) approach \cite{Will1985, Will1987, Konoplya2003, Jerzy2017}, the continued fraction method \cite{Leaver1985}, shooting method \cite{Chandra1975}, etc.\cite{Kono2011, Gund1994, Bin2004}. After scrutinizing these methods, we noticed that the WKB method is probably the most convenient way for our current problem. Therefore, we use the sixth-order WKB method to carry out our calculations.

The formula of $\omega$ from the sixth-order WKB method is given by
\bqn
\lb{WKB1}
\omega &=& \sqrt{-i \left[\left(n+\frac{1}{2}\right)+\sum_{k=2}^6 \Lambda_k \right] \sqrt{-2 V_0''}+V_0},~~~~
\eqn
where
\bqn
\lb{WKB2}
V_0 \equiv \left. V_{\rm eff} \right|_{r=r_{\text{max}}}, \quad V_0'' \equiv \left. \frac{d V_{\rm eff}}{d r^2} \right|_{r=r_{\text{max}}},
\eqn
with {$V_{\rm eff} (r=r_{\text{max}})$} giving the maximum of {$V_{\rm eff}$} on $r \in (1,~\infty)$. The expressions of $\Lambda_k$'s can be found in \cite{Will1985, Will1987, Konoplya2003}. Note that $n=0, 1, 2, ...$.

\begin{table*}
	\caption{The QNMs $\omega$'s  that calculated from the sixth-order WKB method for BHs with DM halos by adopting the background and factors given in case 1 of Table \ref{table0}. Note that each $\omega$ here is compared to its Schwarzschild counterpart.}  
	\label{table1}   
\begin{tabular}{|c c c c c c c c c|} 
		\hline  
		$l$ & \quad  \quad & $n$ & \quad \quad  &  M87 & \quad \quad & Milky Way & \quad \quad  & $\omega$'s of BHs in Schwarzschild case~~
		\\
		\hline
		\hline
		2 & &  0  & &  $0.74718 	-0.17776 i$   & & $ 0.74724 	-0.17778i $ & & $0.74724 	-0.17778i$
		\\  
		  & &   1  & &  $0.69254 	-0.54690 i$    & & $ 0.69259 	-0.54696i $  & &  $0.69259 	-0.54696i$
		\\ 
		  & &  2     & &  $0.59700 	-0.95502 i$   & & $ 0.59704 	-0.95511i $  & &  $0.59704 	-0.95512i$
		\\  
		\hline
		3  & & 0 & &  $1.19879 	-0.18539i$   & & $ 1.19888 	-0.18540i $ & &  $1.19889 	-0.18541i$
		\\  
		   & &   1   & &  $1.16519 	-0.56252i$  & & $ 1.16528 	-0.56258i $  & &  $1.16528 	-0.56258i$
		\\   
		   & &   2   & &  $1.10310 	-0.95799i$   & & $ 1.10318 	-0.95809i $  & &  $1.10319 	-0.95809i$
		\\  
       \hline
		4  & &   0   & &  $1.61823 	-0.18831i$  & & $ 1.61835 	-0.18833i $  & &  $1.61836 	-0.18833i$
		\\  
		   & &   1   & &  $1.59313 	-0.56861i$  & & $ 1.59326 	-0.56866i $  & &  $1.59326 	-0.56867i$
		\\   
		   & &   2   & & $1.54527 	-0.95970 i$   & & $ 1.54538 	-0.95979i $ & &  $1.54539 	-0.95980i$
		\\  
       \hline
	\end{tabular}
\end{table*}

\begin{table*}
	\caption{The QNMs $\omega$'s that calculated from the sixth-order WKB method for BHs with DM halos by adopting the background and factors given in case 2 of Table \ref{table0}. Note that each $\omega$ here is compared to its Schwarzschild counterpart.}  
	\label{table2}   
\begin{tabular}{|c c c c c c c c c|} 
		\hline  
		$l$ & \quad  \quad & $n$ & \quad \quad  &  M87 & \quad \quad & Milky Way & \quad \quad  & $\omega$'s of BHs in Schwarzschild case~~
		\\
		\hline
		\hline
		2 & &  0  & &  $0.74723 	-0.17778 i$   & & $ 0.74723 	-0.17778i $ & & $0.74724 	-0.17778i$
		\\  
		  & &   1  & &   $0.69259 	-0.54695 i$   & & $ 0.69259 	-0.54696i $  & &  $0.69259 	-0.54696i$
		\\ 
		  & &  2     & &  $0.59704 	-0.95511 i$   & & $ 0.59704 	-0.95511i $  & &  $0.59704 	-0.95512i$
		\\  
		\hline
		3  & & 0 & &  $1.19888 	-0.18540i$   & & $ 1.19888 	-0.18540i $ & &  $1.19889 	-0.18541i$
		\\  
		   & &   1   & & $1.16528 	-0.56258 i$   & & $ 1.16528 	-0.56258i $  & &  $1.16528 	-0.56258i$
		\\   
		   & &   2   & &  $1.10318 	-0.95808 i$  & & $ 1.10318 	-0.95809i $  & &  $1.10319 	-0.95809i$
		\\  
       \hline
		4  & &   0   & & $1.61834 	-0.18833 i$   & & $ 1.61835 	-0.18833i $  & &  $1.61836 	-0.18833i$
		\\  
		   & &   1   & & $1.59325 	-0.56866 i$   & & $ 1.59325 	-0.56866i $  & &  $1.59326 	-0.56867i$
		\\   
		   & &   2   & &  $1.54538 	-0.95979 i$  & & $ 1.54538 	-0.95979i $ & &  $1.54539 	-0.95980i$
		\\  
       \hline
	\end{tabular}
\end{table*}

\begin{table*}
\centering
	\caption{The QNMs $\omega$'s that calculated from the sixth-order WKB method for BHs with DM halos by adopting the background and factors given in case 3 of Table \ref{table0}. Note that each $\omega$ here is compared to its Schwarzschild counterpart.}  
	\label{table3}   
\begin{tabular}{|c c c c c c c c c|} 
		\hline  
		$l$ & \quad  \quad & $n$ & \quad \quad  &  M87 & \quad \quad & Milky Way & \quad \quad  & $\omega$'s of BHs in Schwarzschild case~~
		\\
		\hline
		\hline
		2 & &  0  & &    N/A & & $ 0.74724 	-0.17778i $ & & $0.74724 	-0.17778i$
		\\  
		  & &   1  & &    N/A  & & $ 0.69259 	-0.54696i $  & &  $0.69259 	-0.54696i$
		\\ 
		  & &  2     & &  N/A  & & $ 0.59704 	-0.95512i $  & &  $0.59704 	-0.95512i$
		\\  
		\hline
		3  & & 0 & &   N/A  & & $ 1.19888 	-0.18540i $ & &  $1.19889 	-0.18541i$
		\\  
		   & &   1   & &  N/A  & & $ 1.16528 	-0.56258i $  & &  $1.16528 	-0.56258i$
		\\   
		   & &   2   & &  N/A  & & $ 1.10319 	-0.95809i $  & &  $1.10319 	-0.95809i$
		\\  
       \hline
		4  & &   0   & &  N/A  & & $ 1.61835 	-0.18833i $  & &  $1.61836 	-0.18833i$
		\\  
		   & &   1   & &  N/A  & & $ 1.59326 	-0.56867i $  & &  $1.59326 	-0.56867i$
		\\   
		   & &   2   & & N/A  & & $ 1.54539 	-0.95980i $ & &  $1.54539 	-0.95980i$
		\\  
       \hline
	\end{tabular}
\end{table*}

In this subsection, we adopt the choices given by Table \ref{table0} to carry out the calculations. The results of $\omega$'s for this part are exhibited in Tables \ref{table1} - \ref{table3} for case 1, 2 and 3, respectively. Note that in these tables, the results for BHs with DM halos are compared with their Schwarzschild counterparts. It should be mentioned here that during the calculations for case 1, since \eqref{URC_Gr} is quite complicated and will arise technical difficulties when performing derivatives, the {$V_{\rm eff}$} for case 1 is expanded before processing to \eqref{WKB1}. In other words, we use the polynomial form of {$V_{\rm eff}$} in case 1. To be more specific, we expand it as
\bqn
\lb{Veffpoly}
{V_{\rm eff}} &=& \sum^{99}_{k=0} \beta_k \left(r-\frac{79}{50}\right)^k.
\eqn
In this way, we have $\left| \left. {V_{\rm eff} }\right|_{\text{original}} - \left. {V_{\rm eff}} \right|_{\text{polynomial}} \right| \lesssim {\cal{O}} (10^{-19})$ on $r \in (r_{\rm MH},~r_{\text{max}})$ for $l=2,~3,~4$, and thus, the accuracy of QNMs is guaranteed.

By looking at Tables \ref{table1} - \ref{table3}, we immediately notice that the deviations between the Schwarzschild and non-Schwarzschild cases occur at the fourth digit or after that. For most of the $\omega$'s in Tables \ref{table1} - \ref{table3}, these deviations are very small, just like we anticipated earlier. Considering the fact that our calculations contain numerical errors, these deviations are quite negligible. Nonetheless, we can also observe a relatively large deviation on, e.g., $\omega_{40}$ of case 1 for the M87 galaxy, which could probably meet the designed resolution of the TianQin detector {(cf. Table II of \cite{Shi2019})}. That means, according to the current results, we may be able to obtain more constraints for the URC profile by using the observational data from TianQin-like detectors in the future.

\subsection{Tests of the impacts of $\rho_0$ and $r_0$ on QNMs}

In the above subsection, \red{we have considered the impacts of DM halos on QNMs by adopting the halo parameters $\rho_0$ and $r_0$ given in} Table.~\ref{table0}. It is worth mentioning here that these parameters are, in general, derived \red{by fitting the corresponding density profiles with the observational data} of the rotation curves in different galaxies, see \cite{dark_matter} for examples. Thus, these profiles roughly reflect DM distributions for the whole galaxy. They tend to be accurate in describing the regions far away from the central BH. In contrast, to the contexts of the matter environment around the central BH, the halo parameters $\rho_0$ and $r_0$ are basically free. Moreover, in the central region around a BH, the baryonic component also makes significant contributions to the halo parameters. 

\red{In addition, the values of $\rho_0$ and $r_0$ also change from galaxy to galaxy. In the Milky Way, $r_0 \sim 10 \;{\rm kpc}$ and $\rho_0 \sim 10^{7} M_{\odot}/{\rm kpc}^{3}$, as presented in Table.~\ref{table0}. Depending on specific galaxies, it is shown that in the catalog of dark matter halo models for galaxies in the Spitzer Photometry and Accurate Rotation Curves (SPARC) database $r_0$ can be as large as $ \sim 700\; {\rm kpc}$, and $\rho_0$ can be as large as $10^{10} M_{\odot}/{\rm kpc}^{3}$ for a NFW profile \cite{Li:2020iib}. In the SPARC, 175 galaxies are included. Thus, it is natural to expect that the values of $\rho_0$ or $r_0$ in some galaxies in the Universe could be even larger than those given in SPARC.}

For these three reasons, it is worth investigating how the frequencies of QNMs shift with \red{freely changing} $\rho_0$ and $r_0$ in different profiles. We do the calculations based on Table \ref{table0}. Nevertheless, this time we try different values of $\rho_0$ and $r_0$ in addition to those appearing in Table \ref{table0}. Of course, the procedures of calculating QNMs for this part are quite similar to those in the last subsection.

	\begin{figure*}[tbp]
\centering
		\begin{tabular}{cc}
			\includegraphics[width=72mm]{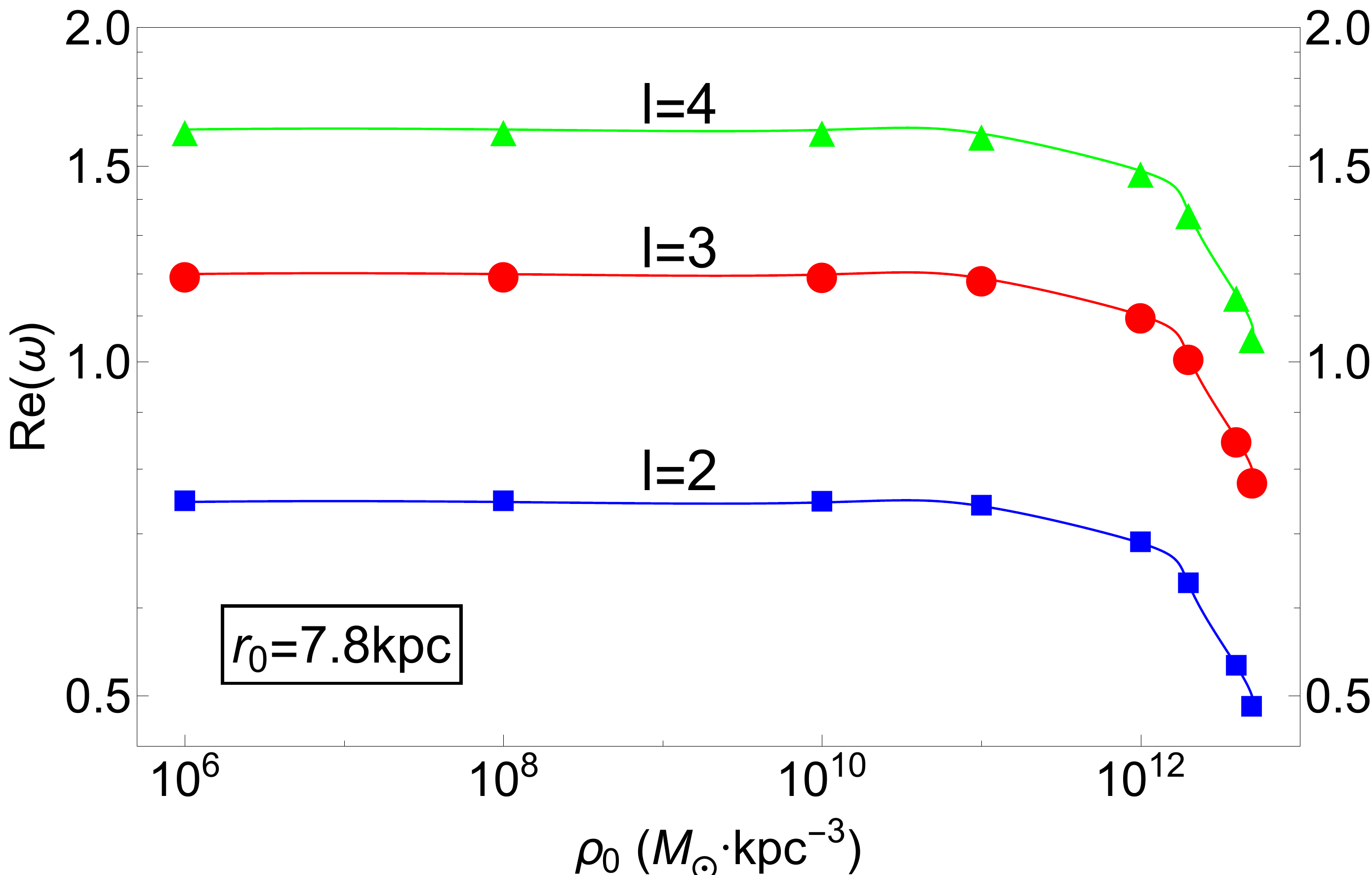} &   \includegraphics[width=78mm]{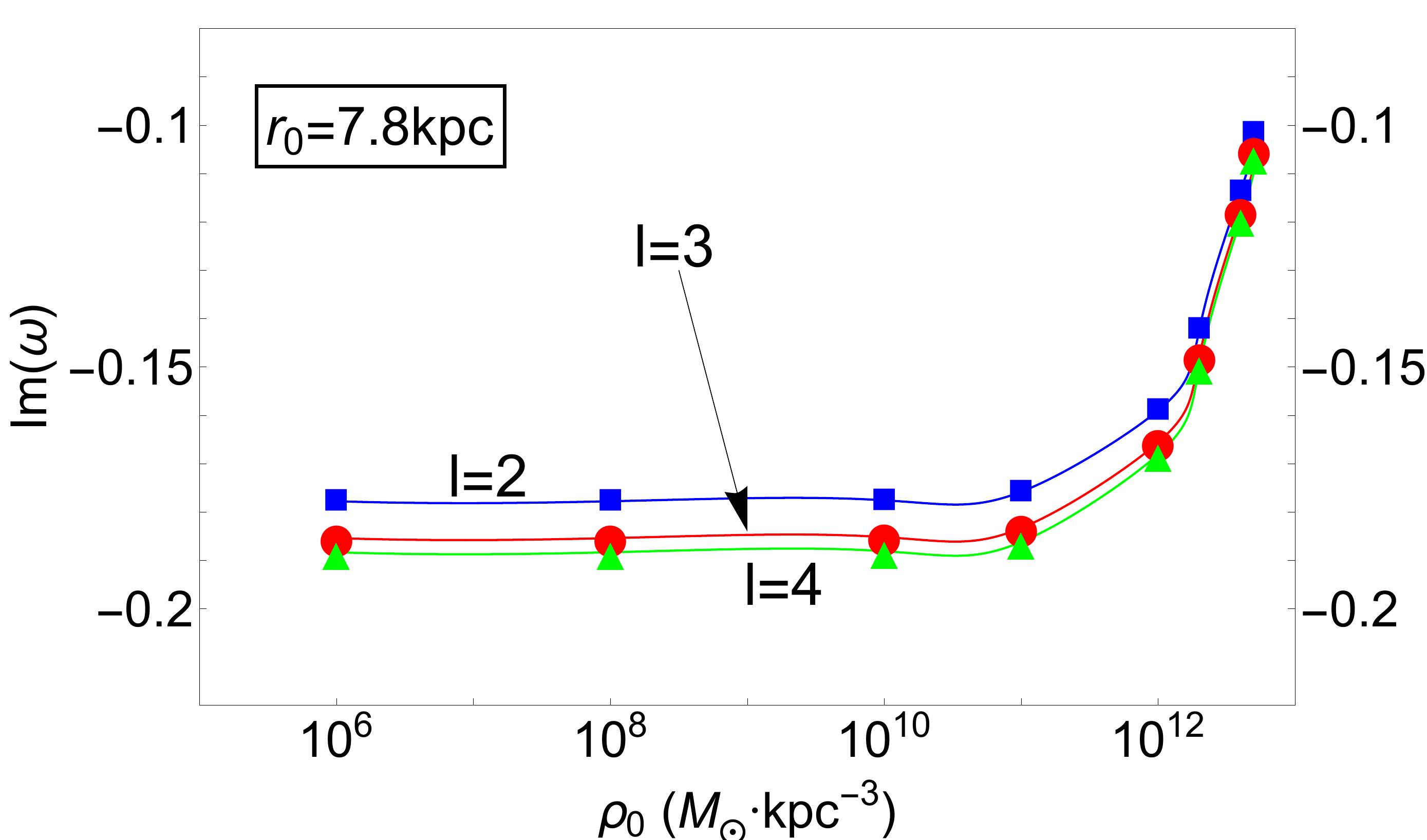} \\
			(a)  & (b)  \\[6pt]
			\includegraphics[width=72mm]{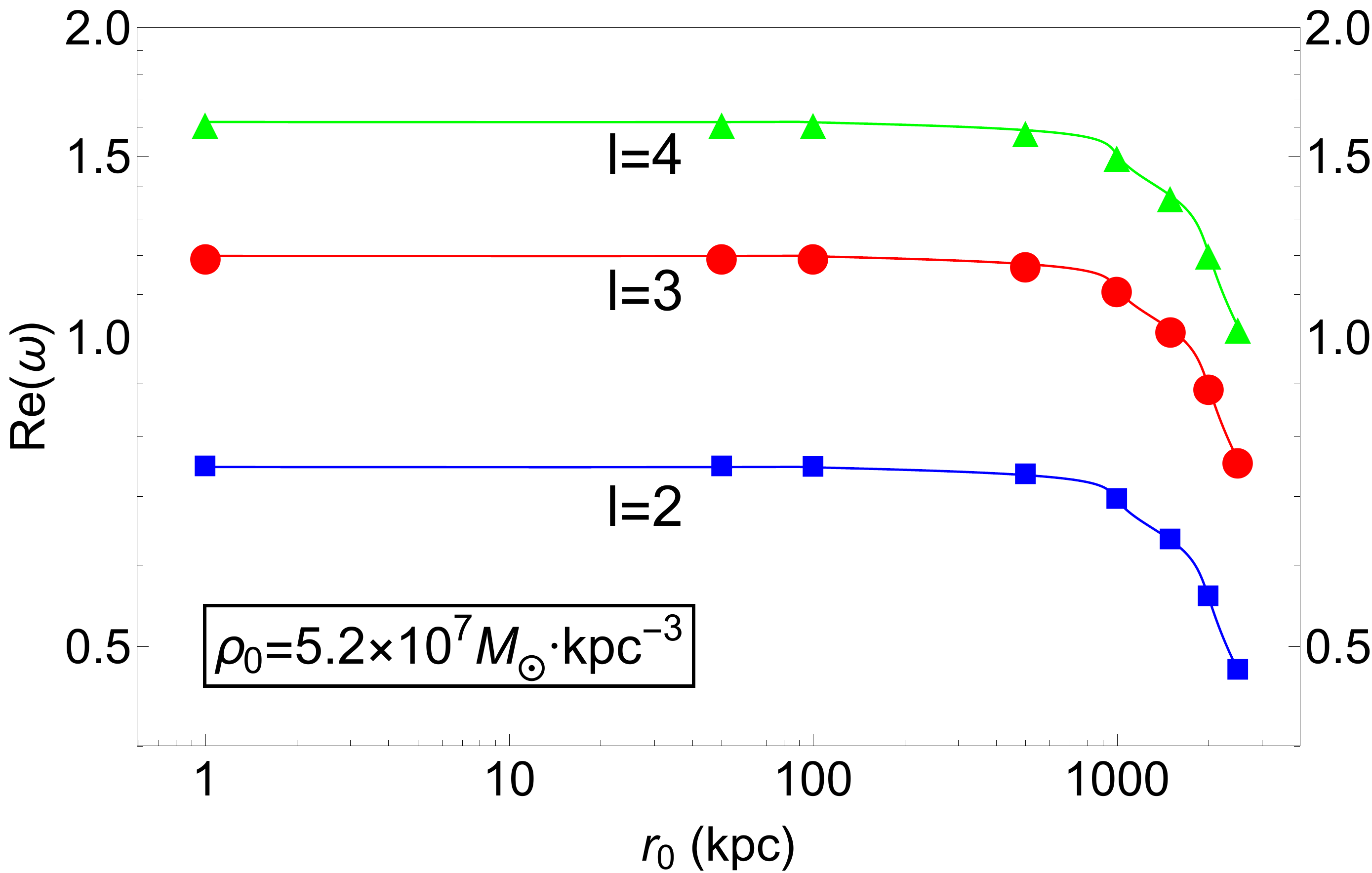} &   \includegraphics[width=78mm]{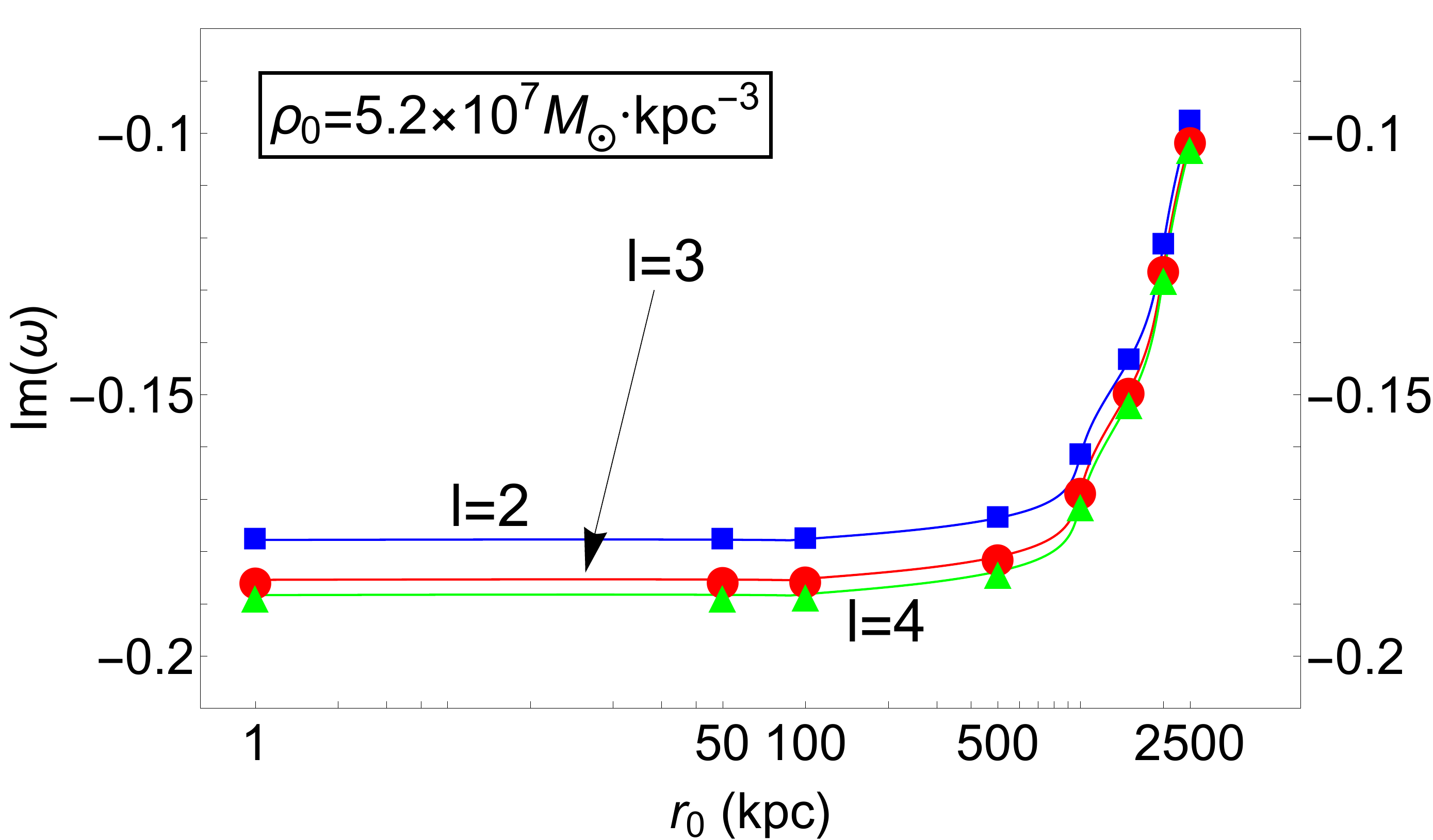} \\
			(c) & (d) \\[6pt]
		\end{tabular}
		\caption{Results of $\omega_{20}$, $\omega_{30}$, and $\omega_{40}$ by changing the values of $\rho_0$ as well as $r_0$ in case 1 (cf. Table \ref{table0}). Panels (a) and (c) show $\text{Re}(\omega)$, while panels (b) and (d)  show $\text{Im}(\omega)$. Note that in panels (a) and (b) $r_0$ is fixed to be 7.8kpc, while in panels (c) and (d) $\rho_0$ is fixed to be $5.2 \times 10^7 M_{\odot}/{\rm kpc}^{3}$. Also note that for the above $\omega$'s we are adopting the unit system so that $c=G_N={r_{\rm MH}}=1$.}
		\label{plot2}
	\end{figure*}

First of all, in Fig. \ref{plot2}, we plot the results of $\omega_{20}$, $\omega_{30}$, and $\omega_{40}$ by changing the values of $\rho_0$ as well as $r_0$ in case 1 (cf. Table \ref{table0}). In panels (a) and (b), the $r_0$ is fixed to be 7.8 kpc, while in panels (c) and (d) the $\rho_0$ is fixed to be $5.2 \times 10^7  M_{\odot}/{\rm kpc}^{3}$. From panels (a) and (b), we observe that both $\text{Re}(\omega)$ and $\text{Im}(\omega)$ are very close to their limits of the Schwarzschild case (cf., e.g,, Table \ref{table1}) at the beginning when $\rho_0 \approx  10^6 M_{\odot}/{\rm kpc}^{3} $. However, these $\omega$'s will soon deviate from the Schwarzschild case when $\rho_0$ is approaching $\rho_0 \approx  10^{12} M_{\odot}/{\rm kpc}^{3} $. Similarly, from panel (c) and (d), we observe that both $\text{Re}(\omega)$ and $\text{Im}(\omega)$ are very close to their limits of the Schwarzschild case at the beginning when $r_0 \approx  1 {\rm kpc} $. However, these $\omega$'s will soon deviate from the Schwarzschild case when $r_0$ is approaching $r_0 \approx 2500 {\rm kpc} $.

\begin{figure*}[tbp]
\centering
		\begin{tabular}{cc}
			\includegraphics[width=72mm]{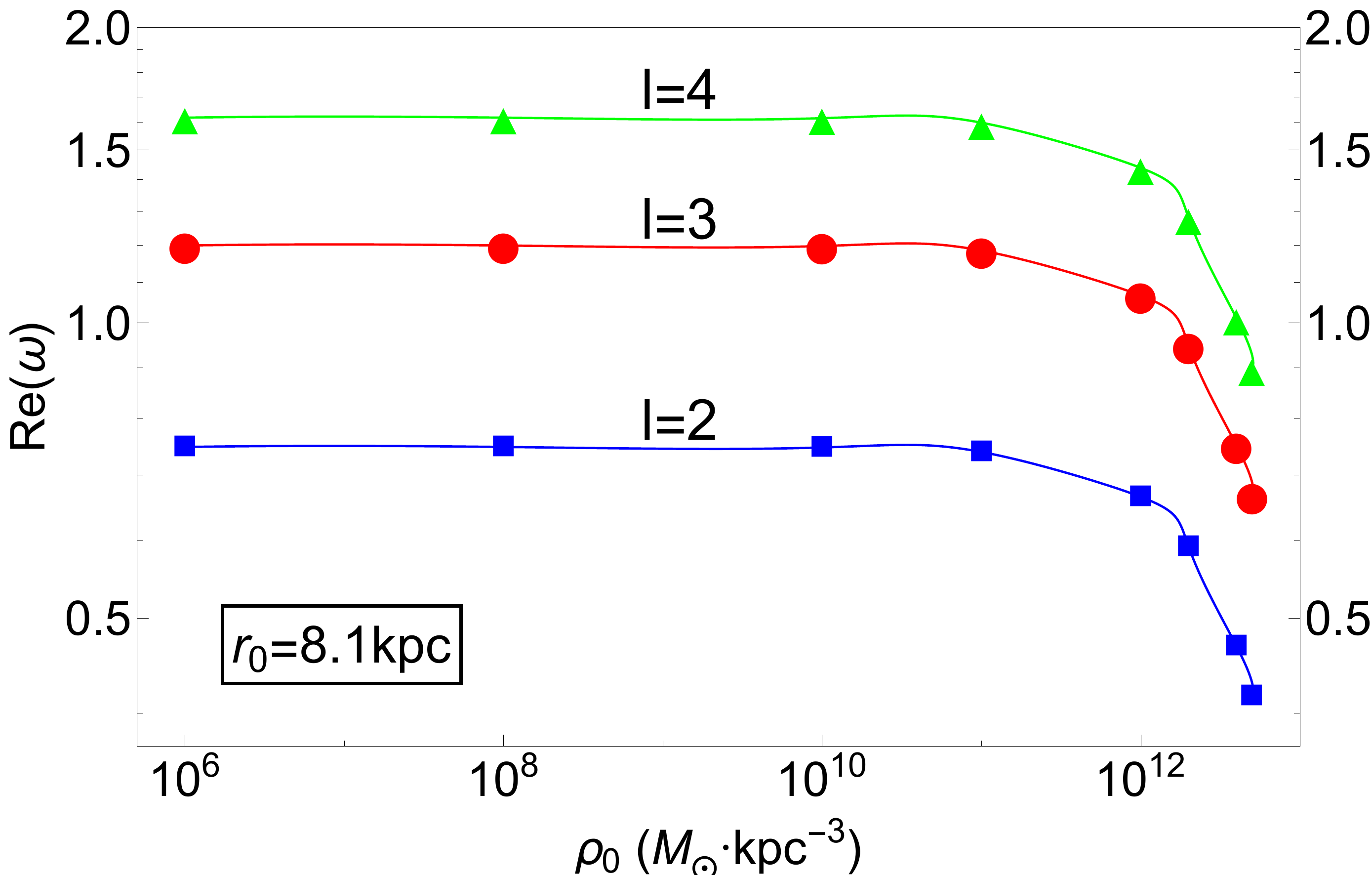} &   \includegraphics[width=78mm]{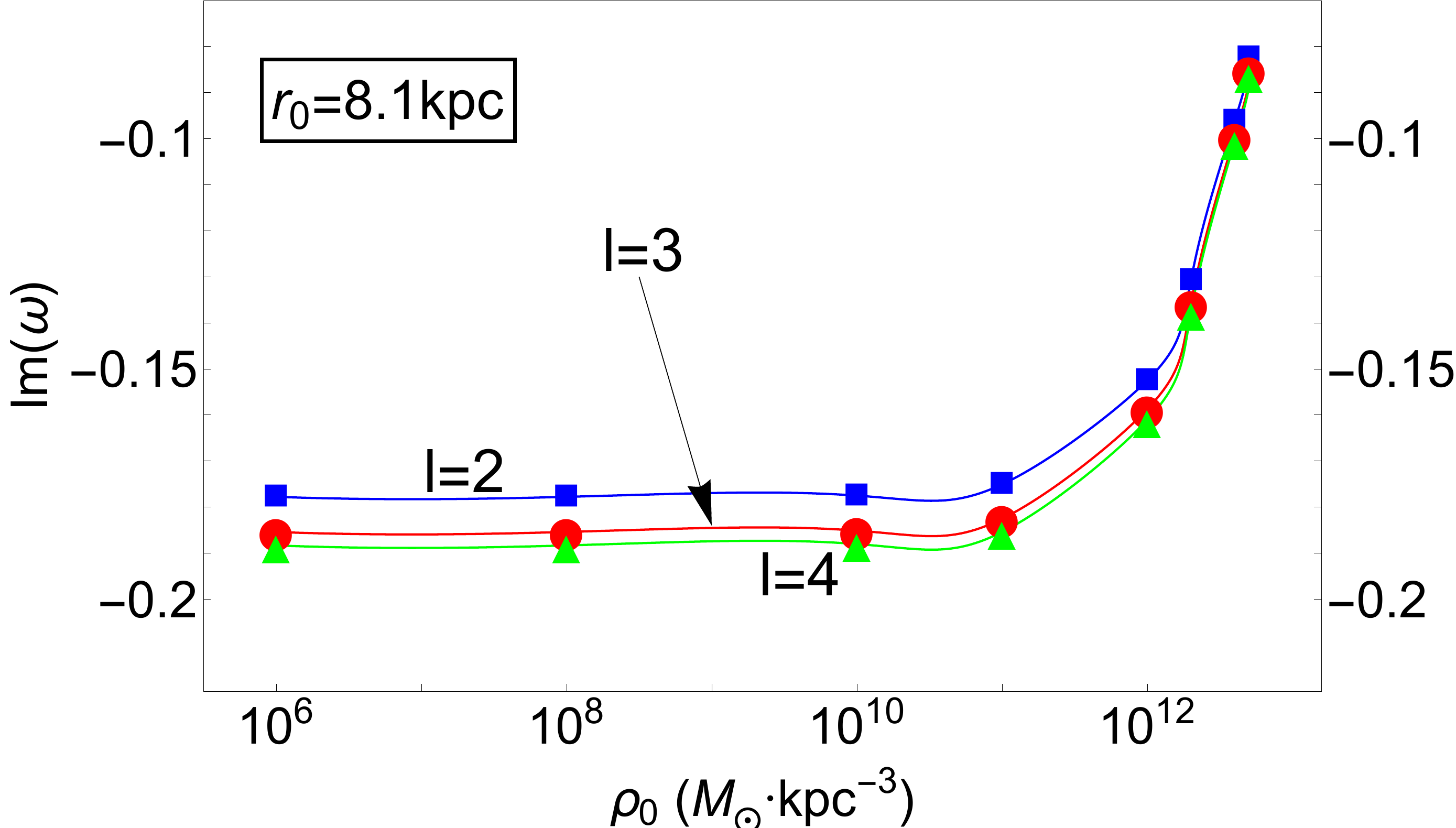} \\
			(a)  & (b)  \\[6pt]
			\includegraphics[width=72mm]{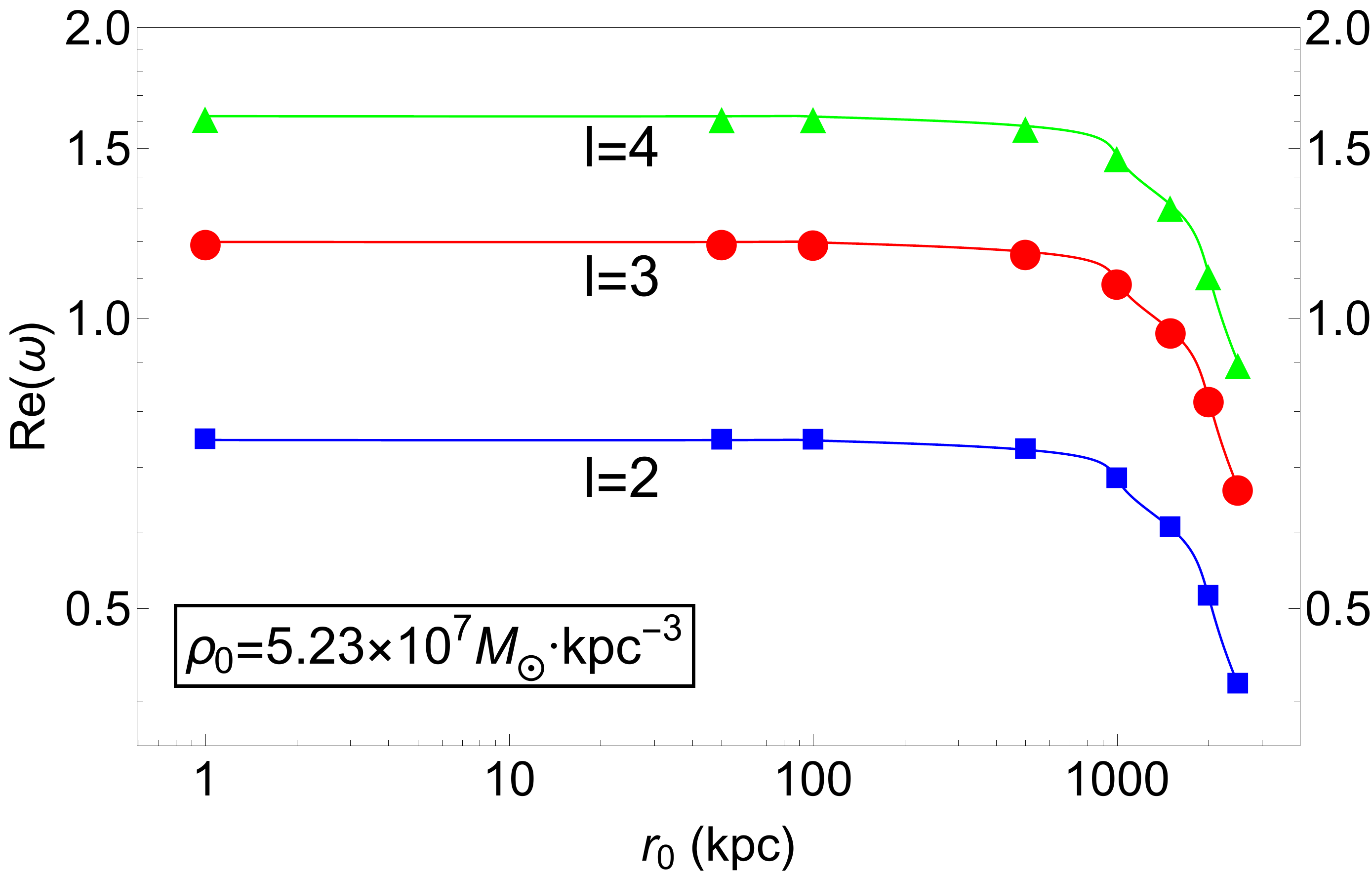} &   \includegraphics[width=78mm]{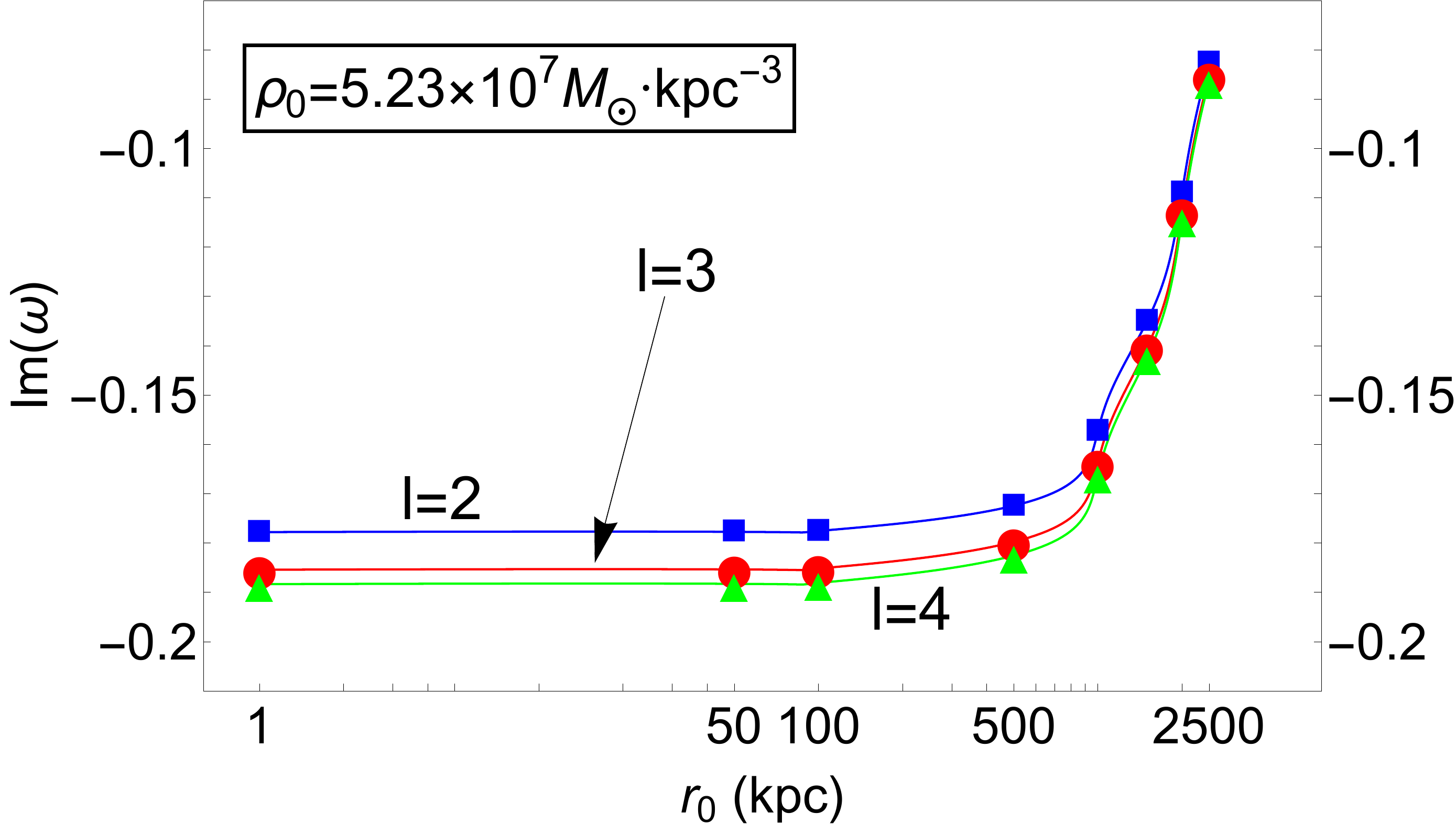} \\
			(c) & (d) \\[6pt]
		\end{tabular}
		\caption{
		Results of $\omega_{20}$, $\omega_{30}$, and $\omega_{40}$ by changing the values of $\rho_0$ as well as $r_0$ in case 2 (cf. Table \ref{table0}). Panels (a) and (c) show $\text{Re}(\omega)$, while panels (b) and (d)  show $\text{Im}(\omega)$. Note that in panels (a) and (b) $r_0$ is fixed to be $8.1 \text{kpc}$, while in panels (c) and (d) $\rho_0$ is fixed to be $5.23 \times 10^7 M_{\odot}/{\rm kpc}^{3}$. Also note that for the above $\omega$'s we are adopting the unit system so that $c=G_N={r_{\rm MH}}=1$.}
		\label{plot3}
	\end{figure*}

\begin{figure*}[tbp]
\centering
		\begin{tabular}{cc}
			\includegraphics[width=72mm]{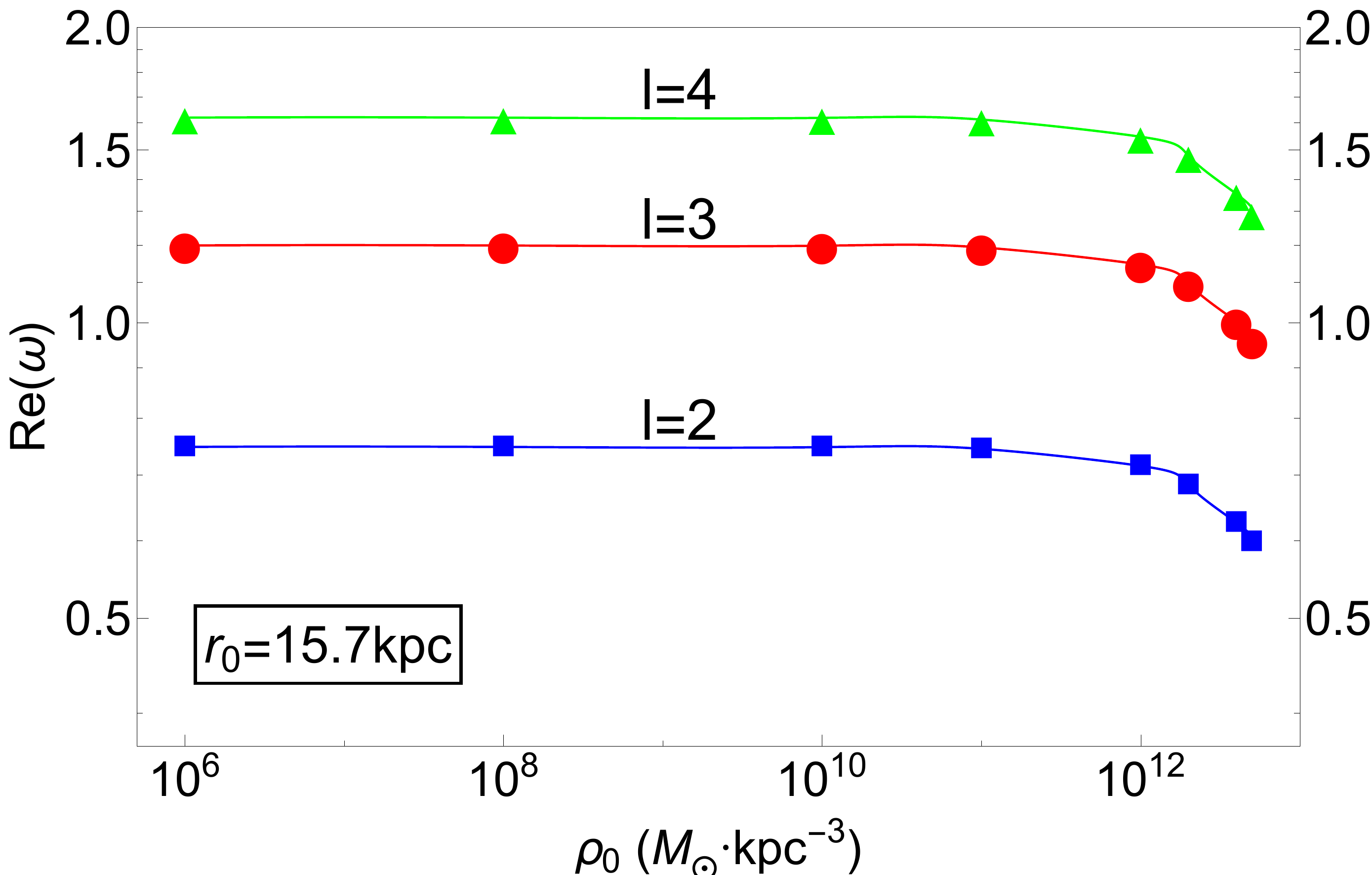} &   \includegraphics[width=78mm]{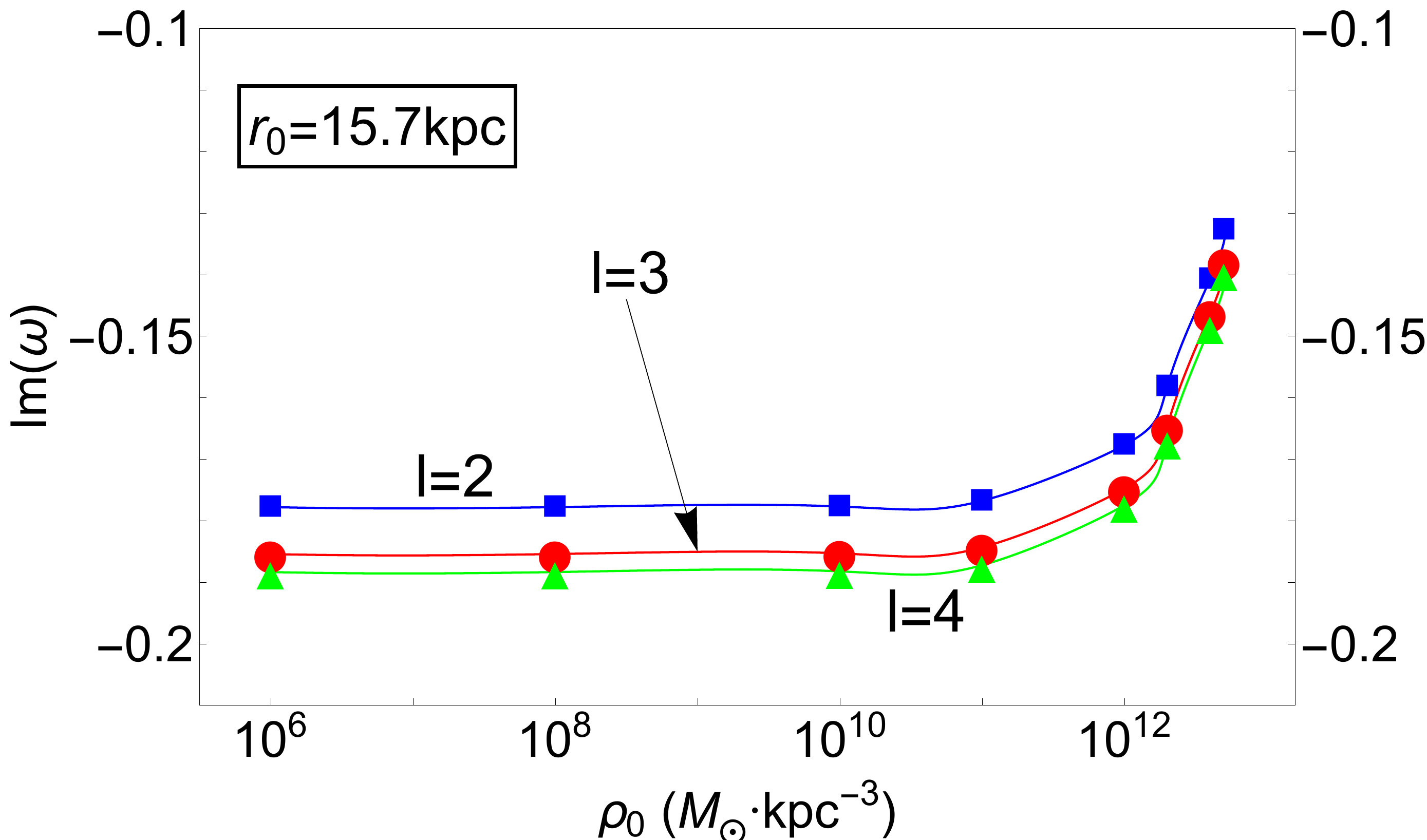} \\
			(a)  & (b)  \\[6pt]
			\includegraphics[width=72mm]{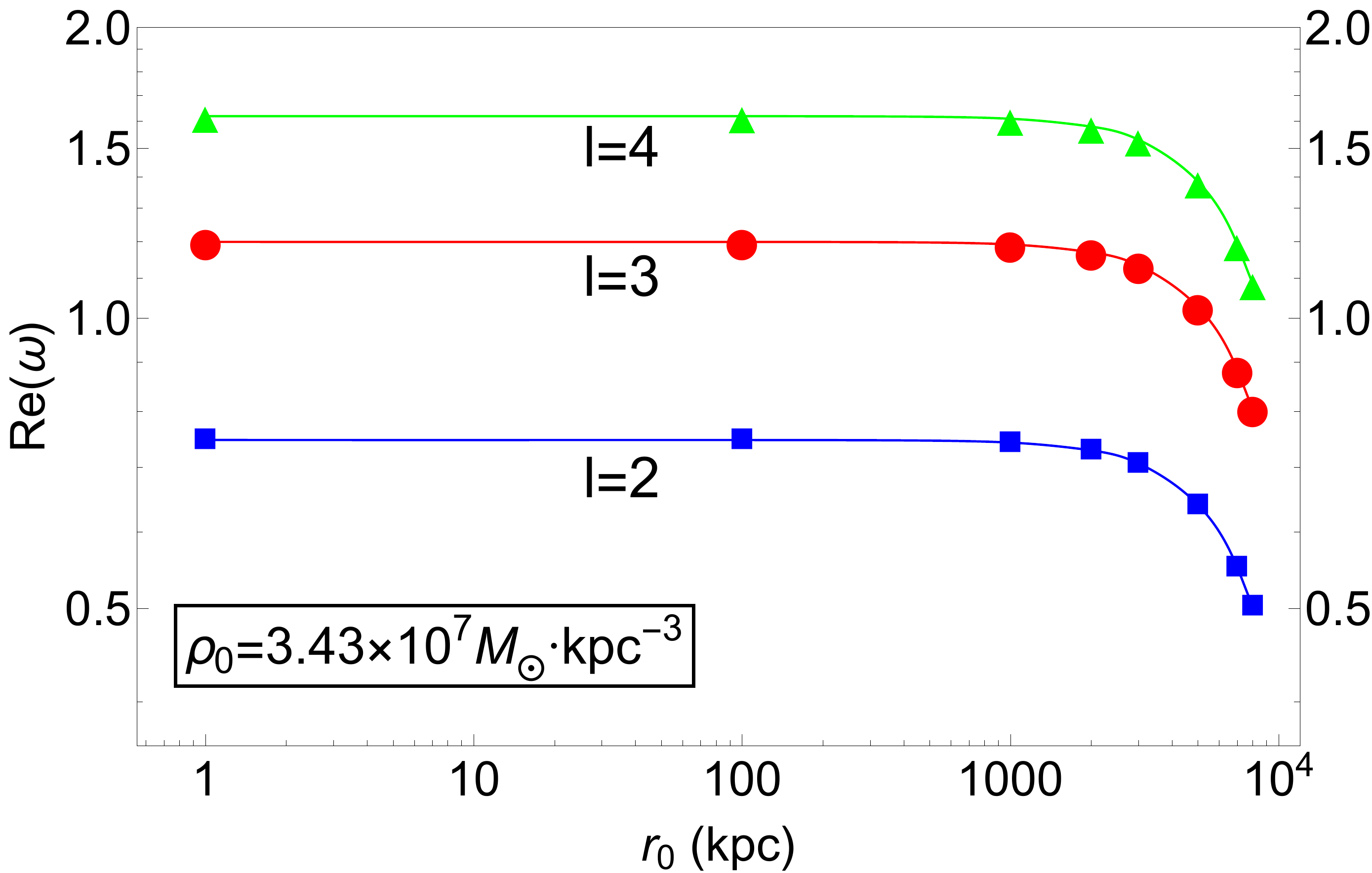} &   \includegraphics[width=78mm]{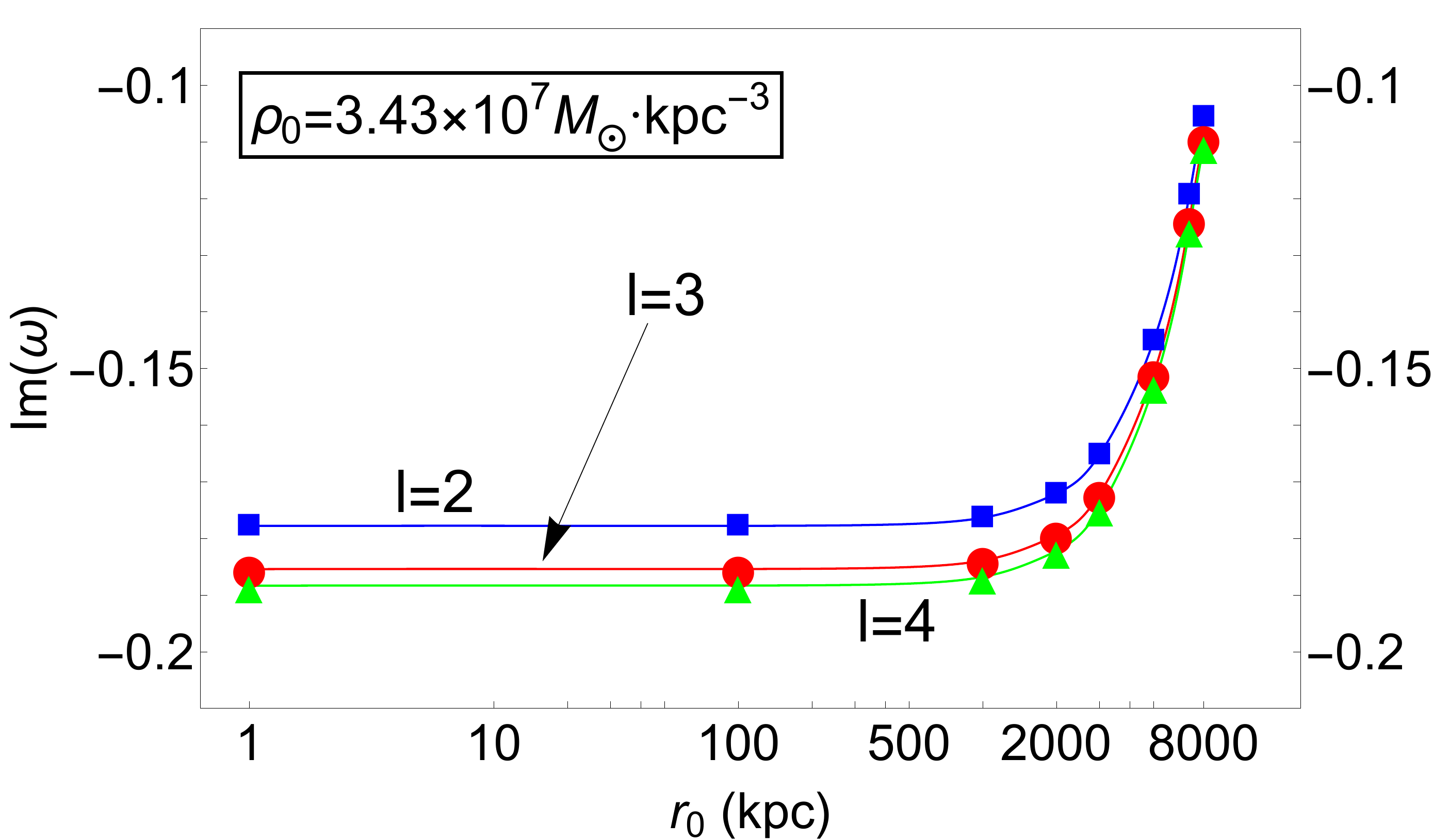} \\
			(c) & (d) \\[6pt]
		\end{tabular}
		\caption{	Results of $\omega_{20}$, $\omega_{30}$ and $\omega_{40}$ by changing the values of $\rho_0$ as well as $r_0$ in Case 3 (cf. Table \ref{table0}). Panel (a) and (c) show $\text{Re}(\omega)$ while Panel (b) and (d)  show $\text{Im}(\omega)$. Note that in Panel (a) and (b) $r_0$ is fixed to be $8.1 \text{kpc}$, while in Panel (c) and (d) $\rho_0$ is fixed to be $3.43 \times 10^7 \text{M}_{\odot}/{\rm kpc}^{3}$. Also note that for the above $\omega$'s we are adopting the unit system so that $c=G_N={r_{\rm MH}}=1$.}
		\label{plot4}
	\end{figure*}

After that, by mimicking Fig. \ref{plot2}, we plot the results of $\omega_{20}$, $\omega_{30}$, and $\omega_{40}$ by changing the values of $\rho_0$ as well as $r_0$ for case 2 and 3 (cf. Table \ref{table0}) in Figs. \ref{plot3} and \ref{plot4}, respectively. Basically, we could see quite similar behaviors of $\omega$'s in these two figures, just like what we have seen in Fig. \ref{plot2}. Therefore, we do not carry out further analysis to these cases.

From Figs.~\ref{plot2}, \ref{plot3}, and \ref{plot4}, it is easy to observe that the QNM frequency is changing as $\rho_0$ or $r_0$ increasing. When $r_0$ is fixed, the real part of the QNM frequency \red{is decreasing,} while the imaginary part \red{is increasing} as $\rho_0$ increases. Similar behaviors also appear when one fixes $\rho_0$ and increases $r_0$. \red{When} $\rho_0$ and $r_0$ become larger, the density of \red{DM becomes bigger accordingly}. These features indicate that the real part of QNM frequency becomes lower, while the imaginary part becomes larger for a denser \red{DM distribution around the BH.} 

In order to see how the above shifts \red{on QNM frequencies affect the GWs from the ringdown stages of coalescences,} we need to construct the corresponding GW waveform. The GW emitted during the ringdown stage can be expressed as a linear combination of damped sinusoids
\bqn
h_{+} + i h_{\times} =\frac{M_z}{D_{\rm L}} \sum_{lmn} {\cal A}_{lmn}\red{e^{i(f_{lmn} t + \phi_{lmn}) } e^{- t/\tau_{lmn}}
}S_{lmn},\nb\\
\eqn
where $M_z$ is the redshifted mass of the black hole, $D_{\rm L}$ the luminosity distance to the source, ${\cal A}_{lmn}$ the mode amplitude, $\phi_{lmn}$ the phase coefficient,  and $S_{lmn}$ the (complex) spin-weighted spheroidal harmonics of spin weight 2, which depend on the polar and azimuthal angles. The GW frequency \red{$f_{lmn} = \text{Re}(\omega_{lmn})$} is the real part of the QNM frequency for $(l, m, n)$ mode, while the damping time $\tau_{lmn}$ is related to the imaginary part of the QNM frequency via $\tau_{lmn} = - 1/\text{Im}(\omega_{lmn})$. From the above waveform, it is evident that a denser DM distribution in the central region near a BH leads to lower GW frequency and longer damping time for \red{GWs} during the ringdown stage. Considering the designed resolution of space-based detectors, such as LISA, TianQin, and Taiji, these effects may be detectable. This may provide an approach to probe the matter distribution in the central region of a galaxy.

\section{Conclusion and Outlooks}
\renewcommand{\theequation}{5.\arabic{equation}} \setcounter{equation}{0}

In this paper, we study the calculations of QNMs for spherically symmetric BHs [cf. \eqref{metric_dark}] with DM halos. We are caring about how DM halos influence the QNMs and thus the GWs from a central BH located in a galaxy. Several different DM profiles are investigated for the M87 galaxy as well as the Milky Way Galaxy, which are referred as case 1, 2, and 3, respectively (cf. Table \ref{table0}). In addition, we have focused on the axial perturbation on the background spacetimes, which are parametrized by \eqref{hab}. 

The backgrounds we consider are described by Eqs.~ \eqref{URC_Gr},  \eqref{Gr2}, and  \eqref{Gr3} for these three different cases. From their expressions we observe that the effects of DM halos are determined by two factors, viz., $r_0$ and $\rho_0$. By using Einstein's field equations [cf. \eqref{Eab}] with a general spherically symmetric background metric [cf. \eqref{backg}], a Regge-Wheeler-like master equation is obtained [cf. \eqref{master2}], for which the RW gauge is adopted \cite{Thomp2017}. With this in hand, and provided a specific set of $\{l, n\}$ (recall that we have set $m=0$), we are able to calculate the corresponding QNMs.

According to the experience with GR, there are many different methods for the calculations of QNMs. Nonetheless, in practice, we noticed that the most convenient way for our problem is the WKB method. Therefore, the sixth-order WKB method [cf. \eqref{WKB1}] is used here for calculating QNMs. The results of $\omega$'s are exhibited in Tables \ref{table1} - \ref{table3} for case 1, 2, and 3 (cf. Table \ref{table0}), respectively. Note that, in there we have selected the unit system so that $c=G_N=r_{\rm MH}=1$.

For most of the $\omega$'s in Tables \ref{table1} - \ref{table3}, we notice that the deviations between the Schwarzschild and non-Schwarzschild  cases are quite small. Considering the fact that our calculations contain numerical errors, these deviations are quite negligible. Nonetheless, we can also observe a relatively large deviation, which occurs at the fourth digit, e.g., $\omega_{40}$ of case 1 for the M87 galaxy. It is worth mentioning here that such a discrepancy could be within the designed resolution of space-based GW detectors, such as LISA, TianQin, and Taiji \cite{Shi2019}. Once they are in operation in the future, one may obtain more constraints for the URC profile by matching the observational data with the results of QNMs in here. More importantly, this result implies the possibility of confining DM profiles with GW observations in the future. 

Since the values of $\rho_0$ and $r_0$ may change from galaxy to galaxy, we also investigate the impacts of them on QNMs. {Indeed, according to the current observations to the Milky Way Galaxy and M87 galaxy, the resultant $\rho_0$ and $r_0$ will lead to quite negligible deviations from the Schwarzschild case (cf., e.g., Fig. \ref{plot1}). Nonetheless, it is hard to tell what kind of parameters we will obtain for other galaxies in the Universe. Thus, it is worth checking what will happen when $\rho_0$ and $r_0$ are changing freely. More importantly, in the context of constraints on BH environments, these parameters are basically free \cite{Cardoso:2021wlq}. It is not necessary to assume that they will preserve similar magnitudes in all the occasions. A more reasonable way is to consider $\rho_0$ and $r_0$ on a wider range.
This fact stimulates our interests on how these parameters will influence the QNMs when they are changing freely.}

Basically, we study the influence of one of the two parameters $\rho_0$ and $r_0$  on QNMs by fixing the other. The corresponding results are shown in Figs. \ref{plot2}-\ref{plot4}. From these figures, it is very clear that the resultant $\omega$'s will approach their limits in the Schwarzschild case when $\rho_0$ and $r_0$ are getting smaller. In contrast, we can see significant deviations on $\omega$'s from the Schwarzschild case for sufficiently large $\rho_0$ and $r_0$. {In addition, we observe that these deviations have been large enough so that we cannot \red{solely} attribute them to numerical errors. For an arbitrary galaxy in the Universe, it is possible to have large enough $\rho_0$ and $r_0$ since they are essentially known to be free parameters. Therefore, according to the results here, a big deviation on QNMs between the Schwarzschild and non-Schwarzschild cases will occur. It is shown that  the real part of QNM frequencies becomes smaller while the imaginary part becomes larger for denser DM distribution in the central region around \red{ a BH.} This implies that \red{the larger dark matter density in the central region near a BH} leads to a lower GW frequency and longer damping time for \red{GWs} during the ringdown stage. Finally, by considering the designed resolution of LISA-like detectors, one may expect such a large deviation to be found in reality once a galaxy with suitable $\rho_0$ and $r_0$ is observed someday. By matching with the results here, these kinds of observations will either confirm our current understanding to DM or help us put constraints on the current DM models.}

Our work here can be extended in several directions. First of all, here, we only consider three different DM profiles. It is interesting to extend the current work to other DM profiles, for instance, BHs surrounded by superfluid DM and baryonic matter \cite{Kimet2020}. On the other hand, since astrophysical BHs, in general, have nonzero angular momentum, it is also our plan to extend our work to rotating BHs. In addition, the even-parity perturbations could also be investigated. Finally, we may test the effects of DM halos on various modified theories of gravity. 

\section*{ACKNOWLEDGEMENTS}

This work is supported by the National Key Research and Development Program of China Grant No. 2020YFC2201503, the Zhejiang Provincial Natural Science Foundation of China under Grants No. LR21A050001 and No. LY20A050002, the National Natural Science Foundation of China under Grants No. 11675143 and No. 11975203, and the Fundamental Research Funds for the Provincial Universities of Zhejiang in China under Grant No. RF-A2019015.

\end{document}